\documentclass[a4paper,11pt]{article}
\usepackage{pos}
\usepackage{dcolumn}
\usepackage{bm}

\def\nuc#1#2#3{{}^{#2}_{#3}\mathrm{#1}}
\def\urm#1{\scriptstyle{\text{\textrm{\textmd{\textup{#1}}}}}}

\def\avr#1{\left\langle{#1}\right\rangle}
\title{QCD sum rule approach to Okamoto-Nolen-Schiffer anomaly}
\author*[a,b]{Hiroyuki Sagawa}
\author[c,d]{Tomoya Naito}
\author[e,f,g]{Xavier Roca-Maza}
\author[c]{Tetsuo Hatsuda}
\affiliation[a]{RIKEN Nishina Center for Accelerator-based Science, Wako 351-0198}
\affiliation[b]{Center for Mathematics and Physics, University of Aizu, 
  Aizu-Wakamatsu, Fukushima 965-8560, Japan}
\affiliation[c]{RIKEN Interdisciplinary Theoretical and Mathematical Sciences Program (iTHEMS),
  Wako 351-0198, Japan}
\affiliation[d]{Department of Physics, Graduate School of Science, The University of Tokyo,
  Tokyo 113-0033, Japan}
\affiliation[e]{Departament de F\'{\i}sica Qu\`{a}ntica i Astrof\'{\i}sica
  and
  Institut de Ci\`{e}ncies del Cosmos,
  Universitat de Barcelona, 
  Mart\'{\i} i Franqu\'{e}s 1,
  08028 Barcelona, Spain}
\affiliation[f]{Dipartimento di Fisica, Universit\`{a} degli Studi di Milano,
  Via Celoria 16, 20133 Milano, Italy}
\affiliation[g]{INFN, Sezione di Milano,
  Via Celoria 16, 20133 Milano, Italy}
\emailAdd{sagawa@ribf.riken.jp}
\abstract{
  A new framework is introduced to connect between a charge symmetry breaking (CSB) energy density functional (EDF) and the low-energy constants derived from quantum chromodynamics (QCD).
  By constructing a QCD-based CSB EDF,
  this method provides new insights into the Okamoto-Nolen-Schiffer anomaly, a long-standing puzzle in the energy differences of mirror nuclei that lacks a robust microscopic explanation.
  Using examples such as
  $ \nuc{F}{17}{} $-$ \nuc{O}{17}{} $,
  $ \nuc{O}{15}{} $-$ \nuc{N}{15}{} $,
  $ \nuc{Sc}{41}{} $-$ \nuc{Ca}{41}{} $, 
  and
  $ \nuc{Ca}{39}{} $-$ \nuc{K}{39}{} $, 
  we demonstrate that the proposed interaction effectively resolves the anomaly within the range of theoretical uncertainties.}
\FullConference{10th International Conference on Quarks and Nuclear Physics (QNP2024) \\
  8--12 July, 2024 \\
  Barcelona, Spain}
\begin{document}
\maketitle
\section{Introduction}
\par
An anomaly in the energy differences of mirror nuclei and isobaric analogue states was found more than 50 years ago and is called
the Okamoto-Nolen-Schiffer (ONS) anomaly~\cite{
  Okamoto1964Phys.Lett.11_150,
  Nolen1969Annu.Rev.Nucl.Sci.19_471}.
This anomaly has been standing many years,  but not yet well understood from a microscopic point of view.  
It was first reported by Okamoto for the case of 
$ \nuc{He}{3}{} $-$ \nuc{H}{3}{} $ system,
and Nolen and Schiffer studied the mass differences of mirror nuclei
systematically from light to heavy nuclei based on the independent-particle model.
They found that the theoretical values of the energy difference underestimate the 
experimental values by $ 3 $--$ 9 \, \% $.
In Table~\ref{tab:table1},
the Coulomb energy differences between mirror nuclei with the mass $ A = 16 \pm 1 $ and $ 40 \pm 1 $ are tabulated.
The Coulomb energies are calculated by Hartree-Fock approximation with a energy density functional (EDF) SGII.
The Coulomb exchange energy is calculated in the exact integration of HF wave functions.
As is clearly seen in Table~\ref{tab:table1},
the calculated Coulomb energy is always smaller than the experimental value by $ 150 $--$ 400 \, \mathrm{keV} $.
\section{ONS anomaly and its physical implications}
\par
To cure this anomaly, 
many extra corrections have been discussed such as
the finite proton size, the center of mass effect, the Thomas-Ehrman effect, 
the isospin impurity, the electromagnetic spin-orbit interaction, the proton-neutron mass difference in the kinetic energy, the core polarization effect, and the
vacuum polarization, altogether explain only about $ 1 \, \% $ of the discrepancy~\cite{
  Shlomo1978Rep.Prog.Phys.41_957},
but found not enough to solve the problem. 
Some possible solutions are further listed as follows:
\begin{enumerate}
\item $ 10 $--$ 20 \, \% $ smaller proton radii of valence particles than those of cores to increase the Coulomb energies of valence particles.
  This prescription will certainly fill the gaps between calculated and experimental energy differences of mirror nuclei.
  However, empirically, the trend is opposite: the proton radii of valence particles 
  are always larger than those of cores.
\item For the charge symmetry breaking interaction (CSB),
  some meson exchange interactions of isospin breaking pairs,
  such as $ \rho $-$ \omega $, $ \pi $-$ \eta $ and $ \delta $-$ \sigma $ meson exchange interactions,
  mediated by photons. 
\item For charge independence breaking interaction (CIB),
  the difference between charged pion and neutral pion exchange interactions.
  The mass difference between two kinds of pions will create CIB interactions in the pion exchange potentials.
\item Introduce phenomenological CSB and CIB interactions such as the Skyrme type.
  The strengths can be  determined by optimizing systematically empirical binding energy differences between mirror nuclei.
\end{enumerate}
\begin{table}[tb]
  \centering
  \caption{Differences of HF Coulomb direct and exchange energies between  mirror nuclei with masses
    $ A = 16 \pm 1 $ and $ 40 \pm 1 $. 
    The SGII EDF is adopted for HF calculations.
    The unit is $ \mathrm{MeV} $.}
  \label{tab:table1}
  \begin{tabular}{lD{.}{.}{3}D{.}{.}{3}D{.}{.}{3}D{.}{.}{3}}
    \hline \hline 
    \multicolumn{1}{l}{Nuclei} & \multicolumn{1}{c}{$ \nuc{F}{17}{} $-$ \nuc{O}{17}{} $} & \multicolumn{1}{c}{$ \nuc{O}{15}{} $-$ \nuc{N}{15}{} $} & \multicolumn{1}{c}{$ \nuc{Sc}{41}{} $-$ \nuc{Ca}{41}{} $} & \multicolumn{1}{c}{$ \nuc{Ca}{39}{} $-$ \nuc{K}{39}{} $} \\
    \hline
    \multicolumn{1}{l}{Orbital} & \multicolumn{1}{c}{$ 1d_{5/2} $} & \multicolumn{1}{c}{$ \left( 1p_{1/2} \right)^{-1} $} & \multicolumn{1}{c}{$ 1f_{7/2} $} & \multicolumn{1}{c}{$ \left( 1d_{3/2} \right)^{-1} $} \\
    \hline
    $ \Delta E_{\rm{D}} $ (Coulomb) &  3.596 & 3.272 &  7.133 & 6.717 \\
    $ \Delta E_{\rm{E}} $ (Coulomb) & -0.203 & 0.026 & -0.267 & 0.260 \\
    \hline
    Sum                              &  3.392 & 3.298 &  6.866 & 6.977 \\ 
    \hline
    Expt.~\cite{Huang2021Chin.Phys.C45_030002} &  3.543 & 3.537 &  7.278 & 7.307 \\
    \hline \hline 
  \end{tabular}
\end{table}
\par
Among them, most plausible source to fill the gap
is the charge symmetry breaking (CSB) nuclear interaction~\cite{
  Okamoto1964Phys.Lett.11_150,
  Negele:1971oux,
  Blunden1987Phys.Lett.B196_295,
  Suzuki1992Nucl.Phys.A536_141,
  Miller:1990iz}.
This idea is originated from the difference of scattering lengths between $ nn $ and $ pp $ scatterings.
The experimental scattering lengths are extracted from the nucleon-nucleon scattering data to be
\begin{equation}
  \label{eq:scat}
  \begin{aligned}
    a_{pp}
    & =
      -17.3 \pm 0.4 \, \mathrm{fm}, \\
    a_{nn}
    & =
      -18.7 \pm 0.6 \, \mathrm{fm}, \\
    a_{pn}
    & =
      -23.7 \pm 0.2 \, \mathrm{fm}.
  \end{aligned}
\end{equation}
The difference between $ a_{pp} $ and $ a_{nn} $ shows the existence of the CSB force,
while the difference between $a_{pn}$ and $ \left( a_{pp} + a_{nn} \right) / 2 $
suggests the CIB force of $ NN $ interaction.
\par 
Recently, 
phenomenological Skyrme-type contact CSB interactions
have been introduced to study the isospin symmetry breaking (ISB) of atomic nuclei systematically
on top of the Coulomb interaction; 
they describe experimental data well, such as  the isobaric analogue states,
the mass differences of iso-doublet and iso-triplet nuclei,
and also the double-beta decays~\cite{
  Roca-Maza2018Phys.Rev.Lett.120_202501,
  Baczyk2018Phys.Lett.B778_178,
  Sagawa2019Eur.Phys.J.A55_227,
  Baczyk2019J.Phys.G46_03LT01,
  Naito2022Phys.Rev.C105_L021304,
  Naito2022Phys.Rev.C106_L061306,
  Naito:2023fnm}.
However, the magnitude and even the sign of the parameters of Skyrme-type CSB interactions have not been well determined.
\textit{Ab initio} calculations for observables sensitive to isospin symmetry breaking terms are also, recently, available~\cite{
  Novario2023Phys.Rev.Lett.130_032501},
while CSB effects have not been isolated in detail and its effects are tend to be much smaller than phenomenological calculation.
\par
The aim of this contribution~\cite{
  QCD-CSB}
is to provide a QCD-based understanding of CSB by making a quantitative link between the Skyrme-type CSB interactions~\cite{
  Sagawa2019Eur.Phys.J.A55_227}
and
the CSB effect due to the $ u $-$ d $ quark mass difference and the associated spontaneous symmetry breaking (SSB) in QCD~\cite{
  Henley1989Phys.Rev.Lett.62_2586,
  Hatsuda1991Phys.Rev.Lett.66_2851,
  Saito:1994tq}. 
To this end, we adopt a QCD sum rule approach, which can provide a microscopic model to discuss the CSB effect based on the idea of 
partial restoration of SSB in the nuclear medium.   
\par
To study QCD dynamics, 
 LQCD is the most fundamental approach, but demands  extremely high computer power and only feasible by usage of top-level super computer.
To avoid such high computational burden, the chiral effective theory and the relativistic mean field model were proposed and applied for nuclear many-body problems successfully.
Analogous theories in nuclear physics and 
condensed matter physics are Hartree-Fock theory and BCS model for superconductor.
QCD sum rule approach has been known an effective and quantitative model to derive the baryon masses and hadron masses in term of the SSB due to the $ q \bar{q} $ condensation in vacuum.
This model can be extended to apply for the study of the nuclear medium effect on SSB, which is intimately related with CSB 
effect of the nucleon-nucleon interaction as will be discussed in the following section.
\par
\begin{figure}[tb]
  \centering
  \includegraphics[width=0.9\linewidth]{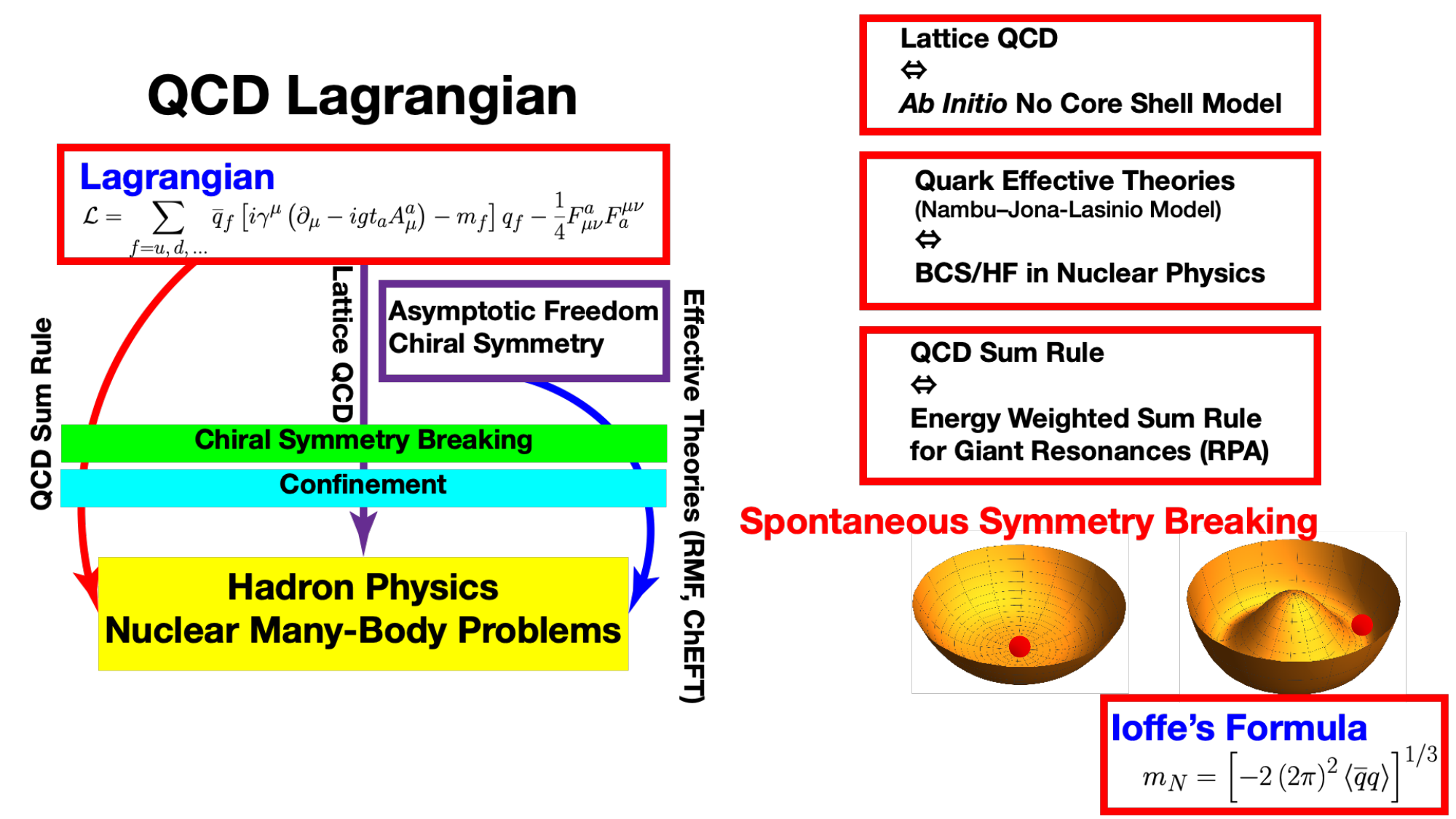}
  \caption{Various QCD approaches for hadron dynamics and quantum many-body problems.  See the text for details.} 
  \label{fig:QCD-Flow}
\end{figure}
\par
\section{CSB EDF from the QCD sum rule and its application}
\par
First, we will discuss the energy difference between the neutron and the proton $ \Delta_{np} \left( \rho \right) $
in symmetric nuclear matter ($ N = Z $) with the baryon density $ \rho $ plus one neutron or proton,
as defined by a difference of the momentum-independent part of the Lorentz-scalar self-energies.
Such difference in the leading order of the $ u $-$ d $ quark mass difference and the quantum electrodynamics (QED) effect
reads
\begin{subequations}
  \label{eq:mass-pn}
  \begin{align}
    \Delta_{np} \left( \rho\right)
    & \simeq 
      C_1 G \left( \rho \right)
      -
      C_2,
      \label{eq:mass-pn_a} \\
    G \left( \rho \right)
    & =
      \left(
      \frac{\langle{\bar{q} q}\rangle}{\langle{\bar{q} q}\rangle_0}
      \right)^{1/3},
      \label{eq:mass-pn_b}
  \end{align}
\end{subequations}
which is approximately obtained from the QCD sum rules (QSR)~\cite{
  Hatsuda1991Phys.Rev.Lett.66_2851}.
Here, $ \avr{\bar{q} q} $ and $ \avr{\bar{q} q}_0 $ are, respectively,
the isospin averaged in-medium and in-vacuum chiral condensate.
The coefficient $ C_1 $ is proportional to  the $ u $-$ d $ quark mass difference 
$ \delta m $
% ~\footnote{
  % The renormalization group invariant mass difference reads
  % $ \delta m \equiv m_d - m_u \simeq 3.6 \, \mathrm{MeV} $~\cite{
  %   FlavourLatticeAveragingGroupFLAG:2021npn}.}, 
through the isospin-breaking constant 
$ \gamma \equiv \avr{\bar{d} d}_0 / \avr{\bar{u} u}_0 - 1 $ 
as $ C_1= - a \gamma $ with a positive numerical constant $ a $ determined by
the Borel QSR method~\cite{
  Hatsuda1991Phys.Rev.Lett.66_2851}.
On the other hand,
$ C_2 $ is a constant originating both from $ \delta m $ and the QED effect.
Equation \eqref{eq:mass-pn} is valid at low density
$ \rho < \rho_0 = 0.17 \, \mathrm{fm}^{-3} $ where the dimension-3 chiral condensate gives a dominant contribution in the operator product expansion in QSR.
In the following, we take
$ C_1 = 5.24^{+2.48}_{-1.21} \, \mathrm{MeV} $.
Since the $ C_2 $-term is density independent, it is cancelled out in the following analysis.
\par
Equation \eqref{eq:mass-pn} implies that $ \Delta_{np} \left( \rho \right) $
decreases due to the partial restoration of the chiral symmetry in medium
[$ G \left( \rho \right) < 1 $].
The in-medium chiral condensate in the leading order with the Fermi-motion correction 
has a universal form~\cite{
  Hayano:2008vn,
  Gubler2019Prog.Part.Nucl.Phys.106_1}
\begin{equation}
  \label{eq:QCD-Q}
  \frac{\langle{\bar{q} q}\rangle}{\langle{\bar{q} q}\rangle_0}
  \simeq
  1
  +
  k_1  \frac{\rho}{\rho_0}
  +
  k_2  \left( \frac{\rho}{\rho_0} \right)^{5/3}
  \quad
  \text{with}
  \quad
  k_1
  =
  -
  \frac{\sigma_{\pi N} \rho_0}{f_{\pi}^2 m_{\pi}^2} < 0,
  \quad
  k_2
  =
  -
  k_1
  \frac{3k_{\urm{F0}}^2}{10 m_N^2}
  >
  0,
\end{equation}
where $ \sigma_{\pi N} $ is the $ \pi $-$ N $ sigma term,
$ m_{\pi} $ ($ m_N $) is the pion (nucleon) mass, and $ f_{\pi}$ is the pion decay constant.
The Fermi-momentum of the symmetric nuclear matter
at saturation is denoted by
$ k_{\urm{F0}} = \left( 3 \pi^2 \rho_0 / 2 \right)^{1/3} = 268 \, \mathrm{MeV} $.
% Systematic calculations using  the in-medium chiral perturbation theory
% shows that the full chiral corrections up to next-to-next-to leading order 
% over Eq.~\eqref{eq:QCD-Q} is numerically 
% small for $ \rho < \rho_0 $~\cite{
%   Goda2013Phys.Rev.C88_065204}.
\par
We decompose the mass difference between mirror nuclei
$ \Delta E = E \left(  Z + 1, N \right) - E \left( Z, N + 1 \right) $
into the Coulomb HF contribution $ \Delta E_{\urm{C}} $ and the ONS anomaly $ \delta_{\urm{ONS}} $ as 
\begin{equation}
  \label{eq:eq11}
  \Delta E
  =
  \Delta E_{\urm{C}}
  +
  \delta_{\urm{ONS}}. 
\end{equation}
Considering Eq.~\eqref{eq:mass-pn},
the CSB effect to $ \delta_{\urm{ONS}} $ estimated based on the QCD sum rule 
from the partial restoration of chiral symmetry in medium reads~\cite{
  Hatsuda1991Phys.Rev.Lett.66_2851}
\begin{equation}
  \label{eq:eq9}
  \delta_{\urm{chiral}}
  \equiv
  \Delta_{np} \left( 0 \right)
  -
  \Delta_{np} \left( \rho \right)
  =
  C_1
  \left[
    1
    - 
    G \left( \rho \right)
  \right].
\end{equation}
\par
Next, we estimate $ \delta_{\urm {ONS}} $ in Eq.~\eqref{eq:eq11} using a Skyrme-type CSB interaction~\cite{
  Sagawa2019Eur.Phys.J.A55_227}
\begin{align}
  V_{\urm {CSB}} \left( \bm{r} \right)
  & =
    \left[
    \vphantom{
    \frac{s_1}{2}
    \left( 1 + y_1 P_{\sigma} \right)
    \left(
    \bm{k}^{\dagger 2} 
    \delta \left( \bm{r} \right)
    +
    \delta \left( \bm{r} \right)
    \bm{k}^2
    \right)}
    s_0 
    \left( 1 + y_0 P_{\sigma} \right)
    \delta \left( \bm{r} \right)
    +
    \frac{s_1}{2}
    \left( 1 + y_1 P_{\sigma} \right)
    \left(
    \bm{k}^{\dagger 2} 
    \delta \left( \bm{r} \right)
    +
    \delta \left( \bm{ r} \right)
    \bm{k}^2
    \right)
    \right.
    \notag \\
  &
    \quad
    \left.
    \vphantom{
    \frac{s_1}{2}
    \left( 1 + y_1 P_{\sigma} \right)
    \left(
    \bm{k}^{\dagger 2} 
    \delta \left( \bm{r} \right)
    +
    \delta \left( \bm{r} \right)
    \bm{k}^2
    \right)}   
    +
    s_2
    \left( 1 + y_2 P_{\sigma} \right)
    \bm{k}^{\dagger}
    \cdot
    \delta \left( \bm{r} \right)
    \bm{k}
    \right]
    \frac{\tau_{1z} + \tau_{2z}}{4},   
    \label{eq:int_CSB}  
\end{align}
where
$ \tau_{iz} = +1 $ ($ -1 $) for neutrons (protons) is
the $ z $-direction of isospin operator of nucleon $ i $,
$ \bm{k} = \left( \bm{ \nabla}_1 - \bm{ \nabla}_2 \right) / 2i $,
$ \bm{r} = \bm{r}_1 - \bm{r}_2 $, 
and $ P_{\sigma} = \left( 1 + \bm{\sigma}_1 \cdot \bm{\sigma}_2 \right) / 2 $
is the spin-exchange operator.
In Eq.~\eqref{eq:int_CSB},
$ s_0 $ and $ y_0 $ are the strength parameters of the contact CSB and its spin-exchange interactions,
while $ s_1 $ ($ s_2 $) and $ y_1 $ ($ y_2 $) are the parameters of the momentum dependent $ s $-wave ($ p $-wave) CSB and its spin-exchange interactions, respectively.  
The general form of $ E \left( Z, N \right) $
for uniform nuclear matter up to the second order of $ \beta = \left( N - Z \right) / A $ reads~\cite{
  Naito:2023fnm}
\begin{equation}
  \label{eq:total-energy}
  \frac{E}{A}
  \simeq
  \varepsilon_0 \left( \rho \right)
  +
  \varepsilon_1 \left( \rho \right)
  \beta
  +
  \varepsilon_2 \left( \rho \right)
  \beta^2 .
\end{equation}
In particular,
we find 
$ \left. \Delta E \right|_{N = Z} = - 2 \varepsilon_1 \left( \rho \right) $ 
where the effect of $ \varepsilon_0 $ and $ \varepsilon_2 $ disappears.
Note that $ \varepsilon_1 $ is non-zero if and only if the CSB interaction exists.
Equation~\eqref{eq:int_CSB} gives contributions to
$ \varepsilon_1 \left( \rho \right) $ and hence $ \delta_{\urm {ONS}} $ as~\cite{
  Naito:2023fnm}
\begin{equation}
  \label{eq:SE-CSB}
  \delta_{\urm {Skyrme}}
  =
  -
  \frac{\tilde{s}_0}{4}
  \rho
  -
  \frac{1}{10}
  \left( \frac{3 \pi^2}{2} \right)^{2/3}
  \left(
    \tilde{s}_1
    +
    3 \tilde{s}_2
  \right)
  \rho^{5/3},
\end{equation}
where we have defined the effective coupling strengths,
\begin{equation}
  \tilde{s}_0
  \equiv
  s_0 \left( 1 - y_0 \right),
  \quad
  \tilde{s}_1
  \equiv
  s_1 \left( 1 - y_1 \right),
  \quad
  \tilde{s}_2
  \equiv
  s_2 \left( 1 + y_2 \right).
\end{equation}
% Note that the Thomas-Fermi approximation is adopted to evaluate the kinetic energy terms in Eq.~\eqref{eq:total-energy}.
\par
\begin{table}[tb]
  \centering
  \caption{Parameters of the Skyrme-type CSB interactions
    constrained from the low-energy constants in QCD.
    To evaluate the CSB effect in finite nuclei where 
    $ \tilde{s}_1 $ and $ \tilde{s}_2 $ contribute independently,
    two characteristic parameter sets (Case I and Case II) are introduced.}
  \label{tab:table2}
  \begin{tabular}{lrr}
    \hline \hline
    $ \tilde{s}_0 $ ($ \mathrm{MeV} \, \mathrm{fm}^3 $)
    & $-15.5^{+8.8}_{-12.5}$ & \\   
    $ \tilde{s}_1 + 3 \tilde{s}_2 $  ($ \mathrm{MeV} \, \mathrm{fm}^5 $)
    & $ 0.52^{+0.42}_{-0.29}$ & \\
    \hline
    & \multicolumn{1}{c}{Case I} & \multicolumn{1}{c}{Case II} \\
    \hline
    $ \tilde{s}_0 $ ($ \mathrm{MeV} \, \mathrm{fm}^3 $) & $ -15.5^{+8.8}_{-12.5} $ & $ -15.5^{+8.8}_{-12.5} $ \\
    $ \tilde{s}_1 $ ($ \mathrm{MeV} \, \mathrm{fm}^5 $) & $ 0.52^{+0.42}_{-0.29} $ & \multicolumn{1}{c}{---}  \\
    $ \tilde{s}_2 $ ($ \mathrm{MeV} \, \mathrm{fm}^5 $) & \multicolumn{1}{c}{---}  & $ 0.18^{+0.14}_{-0.10} $ \\
    \hline \hline
  \end{tabular}
\end{table}
\par
There have been attempts to extract $ \tilde{s}_{\urm{$ 0 $, $ 1 $, $ 2 $}} $ empirically by using various experimental data~\cite{
  Roca-Maza2018Phys.Rev.Lett.120_202501,
  Baczyk2019J.Phys.G46_03LT01}.
These data are summarized in Ref.~\cite{
  Nuovo_Cim}.
Since the contributions of 
$ \tilde{s}_0 $ and $ \tilde{s}_{\urm{$ 1 $, $ 2 $}} $ tend to cancel each other
in physical observables,
it is rather difficult to determine 
the magnitude and the sign
of each term only from the existing experimental data.
\par
On the other hand, our approach is to constrain
$ \tilde{s}_{\urm{$ 0 $, $ 1 $, $ 2 $}} $
from the low-energy constants in QCD,  $ C_1=-a\gamma$ and $ \sigma_{\pi N}$, 
by matching $ \delta_{\urm{Skyrme}} \left( \rho \right) $ in Eq.~\eqref{eq:SE-CSB}
and $ \delta_{\urm{chiral}} \left( \rho \right) $
expanded up to $ \mathcal{O} \left( \rho^{5/3} \right) $ at low densities.
Then, we obtain
\begin{equation}
  \tilde{s}_0
  =
  - \frac{4}{3} 
  \frac{C_1 \sigma_{\pi N}}{f_{\pi}^2 m_{\pi}^2} ,
  \qquad
  \tilde{s}_1 + 3 \tilde{s}_2
  =
  \frac{1}{m_N^2}\frac{C_1 \sigma_{\pi N}}{f_{\pi}^2 m_{\pi}^2}.
\end{equation}
The values $ \tilde{s}_0 $ and $ \tilde{s}_1 + 3 \tilde{s}_2 $ determined are
summarized in Table~\ref{tab:table2}. 
\par
Let us now turn to the comparison of the theoretical values with our CSB interaction
with the experimental mass difference of the mirror nuclei.  
The direct and exchange contributions of the Coulomb interaction
($ \Delta E_{\urm{D}} $ and $ \Delta E_{\urm{E}} $ with
$ \Delta E_{\urm{C}} = \Delta E_{\urm{D}} + \Delta E_{\urm{E}} $)
are obtained with the exact treatment of the exchange term.  
The sum of extra contributions (denoted Extra in Fig. 2) including
the finite-size effect of nucleon,
the center-of-mass effect on nuclear density,
the Thomas-Ehrman effect $ \delta_{\urm{NN}}^1 $,
the isospin impurity $ \delta_{\urm{NN}}^2 $,
the electromagnetic spin-orbit interaction,
the core polarization effect of the last nucleon,
the proton and neutron mass difference in the kinetic energy,
and the vacuum polarization. 
Each contribution varies  from $ -150 \, \mathrm{keV} $ to $ 150 \, \mathrm{keV} $,
while the net result is at most $ 100 \, \mathrm{keV} $ due to a strong cancellation.
As can be seen in Fig. 2, the Extra contributions are too small to fill the difference between $ \Delta E_{\urm{Expt.}} - \Delta E_{\urm{C}} $, but 
the CSB  contributions constrained by the low-energy constants in QCD 
fill the gap quite successfully.
We adopt two EDFs (SGII and SAMi) in the $ \Delta E_{\urm{C}} $ calculations, and found the difference is very small.
Thus, the EDF model dependence is concluded to be a minor effect.   
\begin{figure}[tb]
  \centering
  \includegraphics[width=1.0\linewidth]{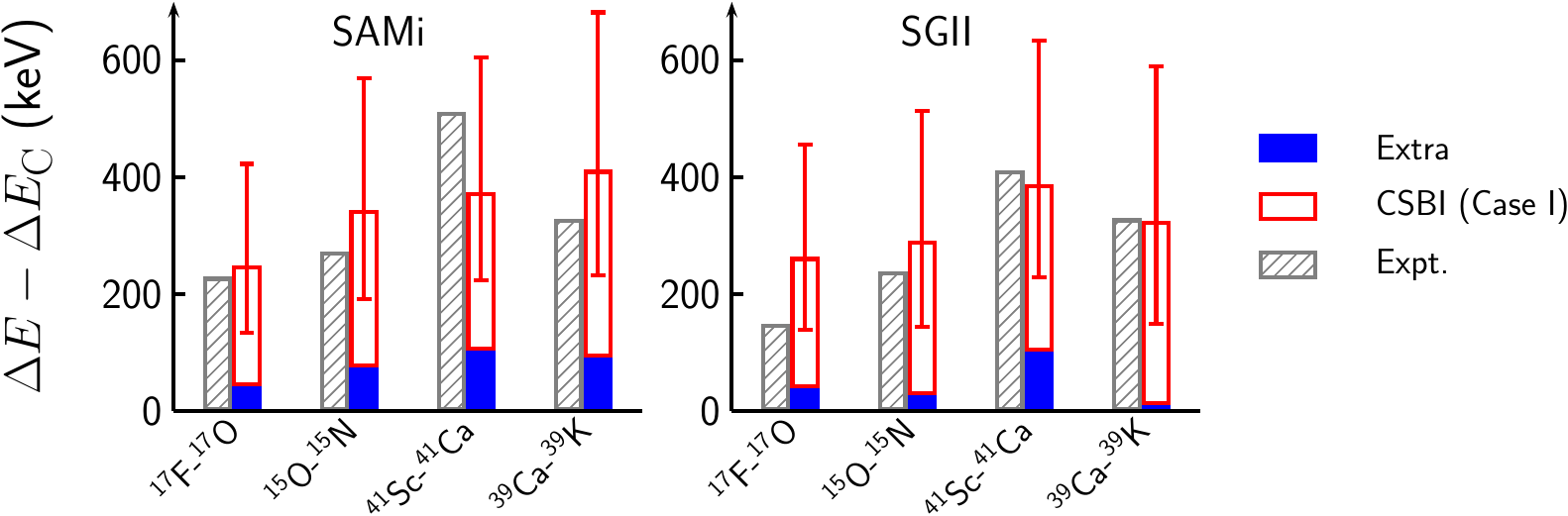}
  \caption{
    Comparisons of the experimental ONS anomaly
    $ \Delta E_{\urm{Expt.}} - \Delta E_{\urm{C}} $ (grey hatched bars)
    and the corresponding theoretical estimates in two EDFs (SGII and SAMi).
    The contribution from the QCD-based CSB interaction (CSBI) in Case I 
    and the extra contributions are
    indicated by the red bars with error bars and the blue bars, respectively.}
  \label{fig:anomaly-crop}
\end{figure}
\section{Summary}
\par
In summary, we determined the EDF parameters of Skyrme-type CSB interactions,
not only the contact term ($ \tilde{s}_0 $)
but also the momentum-dependent terms ($ \tilde{s}_{\urm{$ 1 $, $ 2 $}} $), 
by utilizing the low-energy constants in QCD and the density dependence of chiral condensation of $\bar{q}q$ pair in the nuclear medium.
This is the first attempt to derive the coupling strengths of CSB EDF from the QCD-based approach.
The resulting  CSB interaction is applied to resolve the ONS anomaly:
the numerical results for the mirror nuclei
($ A = 16 \pm 1 $ and $ 40 \pm 1 $ with the isosymmetric core $ N = Z = A/2 $)
with the two Skyrme EDFs (SGII and SAMi)
show good agreement with experimental data both in sign and magnitude within the theoretical uncertainty.
We will further develop a microscopic approach to derive CIB interaction from the mass difference between charged pion and 
neutral pion by the density matrix expansion method,
and apply it for the RPA calculations of IAS in $ N > Z $ nuclei in the next step of the project.

\begin{thebibliography}{29}%
  \makeatletter
  \providecommand \@ifxundefined [1]{%
    \@ifx{#1\undefined}
  }%
  \providecommand \@ifnum [1]{%
    \ifnum #1\expandafter \@firstoftwo
    \else \expandafter \@secondoftwo
    \fi
  }%
  \providecommand \@ifx [1]{%
    \ifx #1\expandafter \@firstoftwo
    \else \expandafter \@secondoftwo
    \fi
  }%
  \providecommand \natexlab [1]{#1}%
  \providecommand \enquote  [1]{``#1''}%
  \providecommand \bibnamefont  [1]{#1}%
  \providecommand \bibfnamefont [1]{#1}%
  \providecommand \citenamefont [1]{#1}%
  \providecommand \href@noop [0]{\@secondoftwo}%
  \providecommand \href [0]{\begingroup \@sanitize@url \@href}%
  \providecommand \@href[1]{\@@startlink{#1}\@@href}%
  \providecommand \@@href[1]{\endgroup#1\@@endlink}%
  \providecommand \@sanitize@url [0]{\catcode `\\12\catcode `\$12\catcode
    `\&12\catcode `\#12\catcode `\^12\catcode `\_12\catcode `\%12\relax}%
  \providecommand \@@startlink[1]{}%
  \providecommand \@@endlink[0]{}%
  \providecommand \url  [0]{\begingroup\@sanitize@url \@url }%
  \providecommand \@url [1]{\endgroup\@href {#1}{\urlprefix }}%
  \providecommand \urlprefix  [0]{URL }%
  \providecommand \Eprint [0]{\href }%
  \providecommand \doibase [0]{https://doi.org/}%
  \providecommand \selectlanguage [0]{\@gobble}%
  \providecommand \bibinfo  [0]{\@secondoftwo}%
  \providecommand \bibfield  [0]{\@secondoftwo}%
  \providecommand \translation [1]{[#1]}%
  \providecommand \BibitemOpen [0]{}%
  \providecommand \bibitemStop [0]{}%
  \providecommand \bibitemNoStop [0]{.\EOS\space}%
  \providecommand \EOS [0]{\spacefactor3000\relax}%
  \providecommand \BibitemShut  [1]{\csname bibitem#1\endcsname}%
  \let\auto@bib@innerbib\@empty
  % </preamble>
\bibitem [{\citenamefont {Okamoto}(1964)}]{Okamoto1964Phys.Lett.11_150}%
  \BibitemOpen
  \bibfield  {author} {\bibinfo {author} {\bibfnamefont {K.}~\bibnamefont
      {Okamoto}},\ }\bibfield  {title} {\bibinfo {title} {{Coulomb energy of $^3$He
        and possible charge asymmetry of nuclear forces}},\ }\href
  {https://doi.org/10.1016/0031-9163(64)90650-X} {\bibfield  {journal}
    {\bibinfo  {journal} {Phys. Lett.}\ }\textbf {\bibinfo {volume} {11}},\
    \bibinfo {pages} {150} (\bibinfo {year} {1964})}\BibitemShut {NoStop}%
\bibitem [{\citenamefont {Nolen}\ and\ \citenamefont
    {Schiffer}(1969)}]{Nolen1969Annu.Rev.Nucl.Sci.19_471}%
  \BibitemOpen
  \bibfield  {author} {\bibinfo {author} {\bibfnamefont {J.~A.}\ \bibnamefont
      {Nolen}, \bibfnamefont {Jr.}}\ and\ \bibinfo {author} {\bibfnamefont {J.~P.}\
      \bibnamefont {Schiffer}},\ }\bibfield  {title} {\bibinfo {title} {{Coulomb
        Energies}},\ }\href {https://doi.org/10.1146/annurev.ns.19.120169.002351}
  {\bibfield  {journal} {\bibinfo  {journal} {Annu. Rev. Nucl. Sci.}\ }\textbf
    {\bibinfo {volume} {19}},\ \bibinfo {pages} {471} (\bibinfo {year}
    {1969})}\BibitemShut {NoStop}%
\bibitem [{\citenamefont {Shlomo}(1978)}]{Shlomo1978Rep.Prog.Phys.41_957}%
  \BibitemOpen
  \bibfield  {author} {\bibinfo {author} {\bibfnamefont {S.}~\bibnamefont
      {Shlomo}},\ }\bibfield  {title} {\bibinfo {title} {{Nuclear Coulomb
        energies}},\ }\href {https://doi.org/10.1088/0034-4885/41/7/001} {\bibfield
    {journal} {\bibinfo  {journal} {Rep. Prog. Phys.}\ }\textbf {\bibinfo
      {volume} {41}},\ \bibinfo {pages} {957} (\bibinfo {year} {1978})}\BibitemShut
  {NoStop}%
\bibitem [{\citenamefont {Negele}(1971)}]{Negele:1971oux}%
  \BibitemOpen
  \bibfield  {author} {\bibinfo {author} {\bibfnamefont {J.~W.}\ \bibnamefont
      {Negele}},\ }\bibfield  {title} {\bibinfo {title} {{The $^{41}$Sc-$^{41}$Ca
        Coulomb energy difference}},\ }\href
  {https://doi.org/10.1016/0375-9474(71)90763-9} {\bibfield  {journal}
    {\bibinfo  {journal} {Nucl. Phys. A}\ }\textbf {\bibinfo {volume} {165}},\
    \bibinfo {pages} {305} (\bibinfo {year} {1971})}\BibitemShut {NoStop}%
\bibitem [{\citenamefont {Blunden}\ and\ \citenamefont
    {Iqbal}(1987)}]{Blunden1987Phys.Lett.B196_295}%
  \BibitemOpen
  \bibfield  {author} {\bibinfo {author} {\bibfnamefont {P.~G.}\ \bibnamefont
      {Blunden}}\ and\ \bibinfo {author} {\bibfnamefont {M.~J.}\ \bibnamefont
      {Iqbal}},\ }\bibfield  {title} {\bibinfo {title} {{Relativistic Hartree-Fock
        calculations for finite nuclei}},\ }\href
  {https://doi.org/10.1016/0370-2693(87)90734-9} {\bibfield  {journal}
    {\bibinfo  {journal} {Phys. Lett. B}\ }\textbf {\bibinfo {volume} {196}},\
    \bibinfo {pages} {295} (\bibinfo {year} {1987})}\BibitemShut {NoStop}%
\bibitem [{\citenamefont {Suzuki}\ \emph {et~al.}(1992)\citenamefont {Suzuki},
    \citenamefont {Sagawa},\ and\ \citenamefont
    {Arima}}]{Suzuki1992Nucl.Phys.A536_141}%
  \BibitemOpen
  \bibfield  {author} {\bibinfo {author} {\bibfnamefont {T.}~\bibnamefont
      {Suzuki}}, \bibinfo {author} {\bibfnamefont {H.}~\bibnamefont {Sagawa}},\
    and\ \bibinfo {author} {\bibfnamefont {A.}~\bibnamefont {Arima}},\ }\bibfield
  {title} {\bibinfo {title} {{Effects of valence nucleon orbits and charge
        symmetry breaking interaction on the Nolen-Schiffer anomally of mirror
        nuclei}},\ }\href {https://doi.org/10.1016/0375-9474(92)90250-N} {\bibfield
    {journal} {\bibinfo  {journal} {Nucl. Phys. A}\ }\textbf {\bibinfo {volume}
      {536}},\ \bibinfo {pages} {141} (\bibinfo {year} {1992})}\BibitemShut
  {NoStop}%
\bibitem [{\citenamefont {Miller}\ \emph {et~al.}(1990)\citenamefont {Miller},
    \citenamefont {Nefkens},\ and\ \citenamefont {Slaus}}]{Miller:1990iz}%
  \BibitemOpen
  \bibfield  {author} {\bibinfo {author} {\bibfnamefont {G.~A.}\ \bibnamefont
      {Miller}}, \bibinfo {author} {\bibfnamefont {B.~M.~K.}\ \bibnamefont
      {Nefkens}},\ and\ \bibinfo {author} {\bibfnamefont {I.}~\bibnamefont
      {Slaus}},\ }\bibfield  {title} {\bibinfo {title} {{Charge symmetry, quarks
        and mesons}},\ }\href {https://doi.org/10.1016/0370-1573(90)90102-8}
  {\bibfield  {journal} {\bibinfo  {journal} {Phys. Rep.}\ }\textbf {\bibinfo
      {volume} {194}},\ \bibinfo {pages} {1} (\bibinfo {year} {1990})}\BibitemShut
  {NoStop}%
\bibitem [{\citenamefont {Roca-Maza}\ \emph {et~al.}(2018)\citenamefont
    {Roca-Maza}, \citenamefont {Col\`{o}},\ and\ \citenamefont
    {Sagawa}}]{Roca-Maza2018Phys.Rev.Lett.120_202501}%
  \BibitemOpen
  \bibfield  {author} {\bibinfo {author} {\bibfnamefont {X.}~\bibnamefont
      {Roca-Maza}}, \bibinfo {author} {\bibfnamefont {G.}~\bibnamefont
      {Col\`{o}}},\ and\ \bibinfo {author} {\bibfnamefont {H.}~\bibnamefont
      {Sagawa}},\ }\bibfield  {title} {\bibinfo {title} {{Nuclear Symmetry Energy
        and the Breaking of the Isospin Symmetry: How Do They Reconcile with Each
        Other?}},\ }\href {https://doi.org/10.1103/PhysRevLett.120.202501} {\bibfield
    {journal} {\bibinfo  {journal} {Phys. Rev. Lett.}\ }\textbf {\bibinfo
      {volume} {120}},\ \bibinfo {pages} {202501} (\bibinfo {year}
    {2018})}\BibitemShut {NoStop}%
\bibitem [{\citenamefont {B\k{a}czyk}\ \emph {et~al.}(2018)\citenamefont
    {B\k{a}czyk}, \citenamefont {Dobaczewski}, \citenamefont {Konieczka},
    \citenamefont {Satu\l{}a}, \citenamefont {Nakatsukasa},\ and\ \citenamefont
    {Sato}}]{Baczyk2018Phys.Lett.B778_178}%
  \BibitemOpen
  \bibfield  {author} {\bibinfo {author} {\bibfnamefont {P.}~\bibnamefont
      {B\k{a}czyk}}, \bibinfo {author} {\bibfnamefont {J.}~\bibnamefont
      {Dobaczewski}}, \bibinfo {author} {\bibfnamefont {M.}~\bibnamefont
      {Konieczka}}, \bibinfo {author} {\bibfnamefont {W.}~\bibnamefont
      {Satu\l{}a}}, \bibinfo {author} {\bibfnamefont {T.}~\bibnamefont
      {Nakatsukasa}},\ and\ \bibinfo {author} {\bibfnamefont {K.}~\bibnamefont
      {Sato}},\ }\bibfield  {title} {\bibinfo {title} {{Isospin-symmetry breaking
        in masses of $ N \simeq Z $ nuclei}},\ }\href
  {https://doi.org/10.1016/j.physletb.2017.12.068} {\bibfield  {journal}
    {\bibinfo  {journal} {Phys. Lett. B}\ }\textbf {\bibinfo {volume} {778}},\
    \bibinfo {pages} {178} (\bibinfo {year} {2018})}\BibitemShut {NoStop}%
\bibitem [{\citenamefont {Sagawa}\ \emph {et~al.}(2019)\citenamefont {Sagawa},
    \citenamefont {Col{\`o}}, \citenamefont {Roca-Maza},\ and\ \citenamefont
    {Niu}}]{Sagawa2019Eur.Phys.J.A55_227}%
  \BibitemOpen
  \bibfield  {author} {\bibinfo {author} {\bibfnamefont {H.}~\bibnamefont
      {Sagawa}}, \bibinfo {author} {\bibfnamefont {G.}~\bibnamefont {Col{\`o}}},
    \bibinfo {author} {\bibfnamefont {X.}~\bibnamefont {Roca-Maza}},\ and\
    \bibinfo {author} {\bibfnamefont {Y.}~\bibnamefont {Niu}},\ }\bibfield
  {title} {\bibinfo {title} {{Collective excitations involving spin and isospin
        degrees of freedom}},\ }\href {https://doi.org/10.1140/epja/i2019-12923-y}
  {\bibfield  {journal} {\bibinfo  {journal} {Eur. Phys. J. A}\ }\textbf
    {\bibinfo {volume} {55}},\ \bibinfo {pages} {227} (\bibinfo {year}
    {2019})}\BibitemShut {NoStop}%
\bibitem [{\citenamefont {B{\k{a}}czyk}\ \emph {et~al.}(2019)\citenamefont
    {B{\k{a}}czyk}, \citenamefont {Satu{\l}a}, \citenamefont {Dobaczewski},\ and\
    \citenamefont {Konieczka}}]{Baczyk2019J.Phys.G46_03LT01}%
  \BibitemOpen
  \bibfield  {author} {\bibinfo {author} {\bibfnamefont {P.}~\bibnamefont
      {B{\k{a}}czyk}}, \bibinfo {author} {\bibfnamefont {W.}~\bibnamefont
      {Satu{\l}a}}, \bibinfo {author} {\bibfnamefont {J.}~\bibnamefont
      {Dobaczewski}},\ and\ \bibinfo {author} {\bibfnamefont {M.}~\bibnamefont
      {Konieczka}},\ }\bibfield  {title} {\bibinfo {title} {{Isobaric multiplet
        mass equation within nuclear density functional theory}},\ }\href
  {https://doi.org/10.1088/1361-6471/aaffe4} {\bibfield  {journal} {\bibinfo
      {journal} {J. Phys. G}\ }\textbf {\bibinfo {volume} {46}},\ \bibinfo {pages}
    {03LT01} (\bibinfo {year} {2019})}\BibitemShut {NoStop}%
\bibitem [{\citenamefont {Naito}\ \emph
    {et~al.}(2022{\natexlab{a}})\citenamefont {Naito}, \citenamefont {Col\`o},
    \citenamefont {Liang}, \citenamefont {Roca-Maza},\ and\ \citenamefont
    {Sagawa}}]{Naito2022Phys.Rev.C105_L021304}%
  \BibitemOpen
  \bibfield  {author} {\bibinfo {author} {\bibfnamefont {T.}~\bibnamefont
      {Naito}}, \bibinfo {author} {\bibfnamefont {G.}~\bibnamefont {Col\`o}},
    \bibinfo {author} {\bibfnamefont {H.}~\bibnamefont {Liang}}, \bibinfo
    {author} {\bibfnamefont {X.}~\bibnamefont {Roca-Maza}},\ and\ \bibinfo
    {author} {\bibfnamefont {H.}~\bibnamefont {Sagawa}},\ }\bibfield  {title}
  {\bibinfo {title} {{Toward \textit{ab initio} charge symmetry breaking in
        nuclear energy density functionals}},\ }\href
  {https://doi.org/10.1103/PhysRevC.105.L021304} {\bibfield  {journal}
    {\bibinfo  {journal} {Phys. Rev. C}\ }\textbf {\bibinfo {volume} {105}},\
    \bibinfo {pages} {L021304} (\bibinfo {year}
    {2022}{\natexlab{a}})}\BibitemShut {NoStop}%
\bibitem [{\citenamefont {Naito}\ \emph
    {et~al.}(2022{\natexlab{b}})\citenamefont {Naito}, \citenamefont {Roca-Maza},
    \citenamefont {Col\`o}, \citenamefont {Liang},\ and\ \citenamefont
    {Sagawa}}]{Naito2022Phys.Rev.C106_L061306}%
  \BibitemOpen
  \bibfield  {author} {\bibinfo {author} {\bibfnamefont {T.}~\bibnamefont
      {Naito}}, \bibinfo {author} {\bibfnamefont {X.}~\bibnamefont {Roca-Maza}},
    \bibinfo {author} {\bibfnamefont {G.}~\bibnamefont {Col\`o}}, \bibinfo
    {author} {\bibfnamefont {H.}~\bibnamefont {Liang}},\ and\ \bibinfo {author}
    {\bibfnamefont {H.}~\bibnamefont {Sagawa}},\ }\bibfield  {title} {\bibinfo
    {title} {{Isospin symmetry breaking in the charge radius difference of mirror
        nuclei}},\ }\href {https://doi.org/10.1103/PhysRevC.106.L061306} {\bibfield
    {journal} {\bibinfo  {journal} {Phys. Rev. C}\ }\textbf {\bibinfo {volume}
      {106}},\ \bibinfo {pages} {L061306} (\bibinfo {year}
    {2022}{\natexlab{b}})}\BibitemShut {NoStop}%
\bibitem [{\citenamefont {Naito}\ \emph {et~al.}(2023)\citenamefont {Naito},
    \citenamefont {Col\`o}, \citenamefont {Liang}, \citenamefont {Roca-Maza},\
    and\ \citenamefont {Sagawa}}]{Naito:2023fnm}%
  \BibitemOpen
  \bibfield  {author} {\bibinfo {author} {\bibfnamefont {T.}~\bibnamefont
      {Naito}}, \bibinfo {author} {\bibfnamefont {G.}~\bibnamefont {Col\`o}},
    \bibinfo {author} {\bibfnamefont {H.}~\bibnamefont {Liang}}, \bibinfo
    {author} {\bibfnamefont {X.}~\bibnamefont {Roca-Maza}},\ and\ \bibinfo
    {author} {\bibfnamefont {H.}~\bibnamefont {Sagawa}},\ }\bibfield  {title}
  {\bibinfo {title} {{Effects of Coulomb and isospin symmetry breaking
        interactions on neutron-skin thickness}},\ }\Eprint
  {https://arxiv.org/abs/2302.08421} {arXiv:2302.08421 [nucl-th]}  (\bibinfo
  {year} {2023})\BibitemShut {NoStop}%
\bibitem [{\citenamefont {Novario}\ \emph {et~al.}(2023)\citenamefont
    {Novario}, \citenamefont {Lonardoni}, \citenamefont {Gandolfi}\ and\ \citenamefont
    {Hagen}}]{Novario2023Phys.Rev.Lett.130_032501}%
  \BibitemOpen
  \bibfield  {author} {\bibinfo {author} {\bibfnamefont {S.~J.}~\bibnamefont
      {Novario}}, \bibinfo {author} {\bibfnamefont {D.}~\bibnamefont
      {Lonardoni}}, \bibinfo {author} {\bibfnamefont {S.}~\bibnamefont
      {Gandolfi}}\ and\ \bibinfo {author} {\bibfnamefont {G.}~\bibnamefont
      {Hagen}},\ }\bibfield  {title} {\bibinfo {title} {{Trends of Neutron Skins and Radii of Mirror Nuclei from First Principles}},\ }\href {https://doi.org/10.1103/PhysRevLett.130.032501} {\bibfield
    {journal} {\bibinfo  {journal} {Phys. Rev. Lett.}\ }\textbf {\bibinfo
      {volume} {130}},\ \bibinfo {pages} {032501} (\bibinfo {year}
    {2023})}\BibitemShut {NoStop}%  
\bibitem [{\citenamefont {Huang}\ \emph {et~al.}(2021)\citenamefont {Huang},
  \citenamefont {Wang}, \citenamefont {Kondev}, \citenamefont {Audi},\ and\
  \citenamefont {Naimi}}]{Huang2021Chin.Phys.C45_030002}%
  \BibitemOpen
  \bibfield  {author} {\bibinfo {author} {\bibfnamefont {W.~J.}\ \bibnamefont
  {Huang}}, \bibinfo {author} {\bibfnamefont {M.}~\bibnamefont {Wang}},
  \bibinfo {author} {\bibfnamefont {F.~G.}\ \bibnamefont {Kondev}}, \bibinfo
  {author} {\bibfnamefont {G.}~\bibnamefont {Audi}},\ and\ \bibinfo {author}
  {\bibfnamefont {S.}~\bibnamefont {Naimi}},\ }\bibfield  {title} {\bibinfo
  {title} {{The AME 2020 atomic mass evaluation (I). Evaluation of input data,
  and adjustment procedures}},\ }\href
  {https://doi.org/10.1088/1674-1137/abddb0} {\bibfield  {journal} {\bibinfo
  {journal} {Chin. Phys. C}\ }\textbf {\bibinfo {volume} {45}},\ \bibinfo
  {pages} {030002} (\bibinfo {year} {2021})}\BibitemShut {NoStop}%
\bibitem [{\citenamefont {Sagawa}\ \emph {et~al.}(2024)\citenamefont {Sagawa},
  \citenamefont {Naito}, \citenamefont {Roca-Maza},\ and\
  \citenamefont {Hatsuda}}]{QCD-CSB}%
  \BibitemOpen
  \bibfield  {author} {\bibinfo {author} {\bibfnamefont {H.}\ \bibnamefont
  {Sagawa}}, \bibinfo {author} {\bibfnamefont {T.}~\bibnamefont {Naito}},
  \bibinfo {author} {\bibfnamefont {X.}\ \bibnamefont {Roca-Maza}},\ and\ \bibinfo {author}
  {\bibfnamefont {T.}~\bibnamefont {Hatsuda}},\ }\bibfield  {title} {\bibinfo
  {title} {{QCD-based charge symmetry breaking interaction and the Okamoto-Nolen-Schiffer anomaly}},\ }\href
  {https://doi.org/10.1103/PhysRevC.109.L011302} {\bibfield  {journal} {\bibinfo
  {journal} {Phys. Rev. C}\ }\textbf {\bibinfo {volume} {109}},\ \bibinfo
  {pages} {L011302} (\bibinfo {year} {2024})}\BibitemShut {NoStop}%
\bibitem [{\citenamefont {Henley}\ and\ \citenamefont
    {Krein}(1989)}]{Henley1989Phys.Rev.Lett.62_2586}%
  \BibitemOpen
  \bibfield  {author} {\bibinfo {author} {\bibfnamefont {E.~M.}\ \bibnamefont
      {Henley}}\ and\ \bibinfo {author} {\bibfnamefont {G.}~\bibnamefont {Krein}},\
  }\bibfield  {title} {\bibinfo {title} {{Nambu--Jona-Lasinio model and charge
        independence}},\ }\href {https://doi.org/10.1103/PhysRevLett.62.2586}
  {\bibfield  {journal} {\bibinfo  {journal} {Phys. Rev. Lett.}\ }\textbf
    {\bibinfo {volume} {62}},\ \bibinfo {pages} {2586} (\bibinfo {year}
    {1989})}\BibitemShut {NoStop}%
\bibitem [{\citenamefont {Hatsuda}\ \emph {et~al.}(1991)\citenamefont
    {Hatsuda}, \citenamefont {H\o{}gaasen},\ and\ \citenamefont
    {Prakash}}]{Hatsuda1991Phys.Rev.Lett.66_2851}%
  \BibitemOpen
  \bibfield  {author} {\bibinfo {author} {\bibfnamefont {T.}~\bibnamefont
      {Hatsuda}}, \bibinfo {author} {\bibfnamefont {H.}~\bibnamefont
      {H\o{}gaasen}},\ and\ \bibinfo {author} {\bibfnamefont {M.}~\bibnamefont
      {Prakash}},\ }\bibfield  {title} {\bibinfo {title} {{QCD sum rules in medium
        and the Okamoto-Nolen-Schiffer anomaly}},\ }\href
  {https://doi.org/10.1103/PhysRevLett.66.2851} {\bibfield  {journal} {\bibinfo
      {journal} {Phys. Rev. Lett.}\ }\textbf {\bibinfo {volume} {66}},\ \bibinfo
    {pages} {2851} (\bibinfo {year} {1991})}\BibitemShut {NoStop}%
\bibitem [{\citenamefont {Saito}\ and\ \citenamefont
    {Thomas}(1994)}]{Saito:1994tq}%
  \BibitemOpen
  \bibfield  {author} {\bibinfo {author} {\bibfnamefont {K.}~\bibnamefont
      {Saito}}\ and\ \bibinfo {author} {\bibfnamefont {A.~W.}\ \bibnamefont
      {Thomas}},\ }\bibfield  {title} {\bibinfo {title} {{The Nolen-Schiffer
        anomaly and isospin symmetry breaking in nuclear matter}},\ }\href
  {https://doi.org/10.1016/0370-2693(94)91551-2} {\bibfield  {journal}
    {\bibinfo  {journal} {Phys. Lett. B}\ }\textbf {\bibinfo {volume} {335}},\
    \bibinfo {pages} {17} (\bibinfo {year} {1994})}\BibitemShut {NoStop}%
% \bibitem [{\citenamefont {Aoki}\ \emph {et~al.}(2022)\citenamefont {Aoki} \emph
%   {et~al.}}]{FlavourLatticeAveragingGroupFLAG:2021npn}%
%   \BibitemOpen
%   \bibfield  {author} {\bibinfo {author} {\bibfnamefont {Y.}~\bibnamefont
%   {Aoki}} \emph {et~al.} (\bibinfo {collaboration} {Flavour Lattice Averaging
%   Group (FLAG)}),\ }\bibfield  {title} {\bibinfo {title} {{FLAG Review 2021}},\
%   }\href {https://doi.org/10.1140/epjc/s10052-022-10536-1} {\bibfield
%   {journal} {\bibinfo  {journal} {Eur. Phys. J. C}\ }\textbf {\bibinfo {volume}
%   {82}},\ \bibinfo {pages} {869} (\bibinfo {year} {2022})}\BibitemShut
%   {NoStop}%  
\bibitem [{\citenamefont {Hayano}\ and\ \citenamefont
    {Hatsuda}(2010)}]{Hayano:2008vn}%
  \BibitemOpen
  \bibfield  {author} {\bibinfo {author} {\bibfnamefont {R.~S.}\ \bibnamefont
      {Hayano}}\ and\ \bibinfo {author} {\bibfnamefont {T.}~\bibnamefont
      {Hatsuda}},\ }\bibfield  {title} {\bibinfo {title} {{Hadron properties in the
        nuclear medium}},\ }\href {https://doi.org/10.1103/RevModPhys.82.2949}
  {\bibfield  {journal} {\bibinfo  {journal} {Rev. Mod. Phys.}\ }\textbf
    {\bibinfo {volume} {82}},\ \bibinfo {pages} {2949} (\bibinfo {year}
    {2010})}\BibitemShut {NoStop}%
\bibitem [{\citenamefont {Gubler}\ and\ \citenamefont
    {Satow}(2019)}]{Gubler2019Prog.Part.Nucl.Phys.106_1}%
  \BibitemOpen
  \bibfield  {author} {\bibinfo {author} {\bibfnamefont {P.}~\bibnamefont
      {Gubler}}\ and\ \bibinfo {author} {\bibfnamefont {D.}~\bibnamefont {Satow}},\
  }\bibfield  {title} {\bibinfo {title} {{Recent progress in QCD condensate
        evaluations and sum rules}},\ }\href
  {https://doi.org/10.1016/j.ppnp.2019.02.005} {\bibfield  {journal} {\bibinfo
      {journal} {Prog. Part. Nucl. Phys.}\ }\textbf {\bibinfo {volume} {106}},\
    \bibinfo {pages} {1} (\bibinfo {year} {2019})}\BibitemShut {NoStop}%
\bibitem [{\citenamefont {Naito}\ \emph {et~al.}(2024)\citenamefont
    {Naito}, \citenamefont {Col\`{o}}, \citenamefont {Hatsuda},
    \citenamefont {Liang}, \citenamefont {Roca-Maza},\ and\ \citenamefont
    {Sagawa}}]{Nuovo_Cim}%
  \BibitemOpen
  \bibfield  {author} {\bibinfo {author} {\bibfnamefont {T.}~\bibnamefont
      {Naito}}, \bibinfo {author} {\bibfnamefont {G.}~\bibnamefont
      {Col\`{o}}}, \bibinfo {author} {\bibfnamefont {T.}~\bibnamefont
      {Hatsuda}}, \bibinfo {author} {\bibfnamefont {H.}~\bibnamefont
      {Liang}}, \bibinfo {author} {\bibfnamefont {X.}~\bibnamefont
      {Roca-Maza}}, \ and\ \bibinfo {author} {\bibfnamefont {H.}~\bibnamefont {Sagawa}},\
  }\bibfield  {title} {\bibinfo {title} {{Possible inconsistency between phenomenological and theoretical determinations of charge symmetry breaking in nuclear energy density functionals}},\ }\href
  {https://doi.org/10.1393/ncc/i2024-24052-9} {\bibfield  {journal} {\bibinfo
      {journal} {Nuovo Cim. C}\ }\textbf {\bibinfo {volume} {47}},\
    \bibinfo {pages} {52} (\bibinfo {year} {2024})}\BibitemShut {NoStop}%
% \bibitem [{\citenamefont {Goda}\ and\ \citenamefont
%     {Jido}(2013)}]{Goda2013Phys.Rev.C88_065204}%
%   \BibitemOpen
%   \bibfield  {author} {\bibinfo {author} {\bibfnamefont {S.}~\bibnamefont
%       {Goda}}\ and\ \bibinfo {author} {\bibfnamefont {D.}~\bibnamefont {Jido}},\
%   }\bibfield  {title} {\bibinfo {title} {{Chiral condensate at finite density
%         using the chiral Ward identity}},\ }\href
%   {https://doi.org/10.1103/PhysRevC.88.065204} {\bibfield  {journal} {\bibinfo
%       {journal} {Phys. Rev. C}\ }\textbf {\bibinfo {volume} {88}},\ \bibinfo
%     {pages} {065204} (\bibinfo {year} {2013})}\BibitemShut {NoStop}%
% \bibitem [{\citenamefont {Kaiser}\ \emph {et~al.}(2008)\citenamefont {Kaiser},
%     \citenamefont {de~Homont},\ and\ \citenamefont
%     {Weise}}]{Kaiser2008Phys.Rev.C77_025204}%
%   \BibitemOpen
%   \bibfield  {author} {\bibinfo {author} {\bibfnamefont {N.}~\bibnamefont
%       {Kaiser}}, \bibinfo {author} {\bibfnamefont {P.}~\bibnamefont {de~Homont}},\
%     and\ \bibinfo {author} {\bibfnamefont {W.}~\bibnamefont {Weise}},\ }\bibfield
%   {title} {\bibinfo {title} {{In-medium chiral condensate beyond linear
%         density approximation}},\ }\href {https://doi.org/10.1103/PhysRevC.77.025204}
%   {\bibfield  {journal} {\bibinfo  {journal} {Phys. Rev. C}\ }\textbf {\bibinfo
%       {volume} {77}},\ \bibinfo {pages} {025204} (\bibinfo {year}
%     {2008})}\BibitemShut {NoStop}%
% \bibitem [{\citenamefont {Gasser}\ \emph {et~al.}(1991)\citenamefont {Gasser},
%     \citenamefont {Leutwyler},\ and\ \citenamefont {Sainio}}]{Gasser:1990ce}%
%   \BibitemOpen
%   \bibfield  {author} {\bibinfo {author} {\bibfnamefont {J.}~\bibnamefont
%       {Gasser}}, \bibinfo {author} {\bibfnamefont {H.}~\bibnamefont {Leutwyler}},\
%     and\ \bibinfo {author} {\bibfnamefont {M.~E.}\ \bibnamefont {Sainio}},\
%   }\bibfield  {title} {\bibinfo {title} {{Sigma term update}},\ }\href
%   {https://doi.org/10.1016/0370-2693(91)91393-A} {\bibfield  {journal}
%     {\bibinfo  {journal} {Phys. Lett. B}\ }\textbf {\bibinfo {volume} {253}},\
%     \bibinfo {pages} {252} (\bibinfo {year} {1991})}\BibitemShut {NoStop}%
% \bibitem [{\citenamefont {De~Vries}\ \emph {et~al.}(1987)\citenamefont
%     {De~Vries}, \citenamefont {De~Jager},\ and\ \citenamefont
%     {De~Vries}}]{DeVries1987At.DataNucl.DataTables36_495}%
%   \BibitemOpen
%   \bibfield  {author} {\bibinfo {author} {\bibfnamefont {H.}~\bibnamefont
%       {De~Vries}}, \bibinfo {author} {\bibfnamefont {C.~W.}\ \bibnamefont
%       {De~Jager}},\ and\ \bibinfo {author} {\bibfnamefont {C.}~\bibnamefont
%       {De~Vries}},\ }\bibfield  {title} {\bibinfo {title} {{Nuclear
%         charge-density-distribution parameters from elastic electron scattering}},\
%   }\href {https://doi.org/10.1016/0092-640X(87)90013-1} {\bibfield  {journal}
%     {\bibinfo  {journal} {At. Data Nucl. Data Tables}\ }\textbf {\bibinfo
%       {volume} {36}},\ \bibinfo {pages} {495} (\bibinfo {year} {1987})}\BibitemShut
%   {NoStop}%
\end{thebibliography}
\end{document}